\documentclass[12pt]{article}
\pdfoutput=1

\usepackage[english]{babel}
\usepackage{amssymb,amsmath}
\usepackage{amsfonts}
\usepackage{latexsym}
\usepackage{verbatim}
\usepackage{amsthm}
\usepackage{amsbsy}
\usepackage{graphicx,color}
\usepackage{epsfig}
\usepackage{verbatim}
\usepackage{tcolorbox}
\tcbuselibrary{theorems}
\usepackage[all]{xy}

\usepackage{cite}
\usepackage[debug,pageanchor=false]{hyperref}
\definecolor{link}{rgb}{.8,.15,.1}
\hypersetup{colorlinks=true,linkcolor=link,citecolor=link,urlcolor=link,linktocpage}

\newcommand{\be}{\begin{equation}}
\newcommand{\ee}{\end{equation}}
\newcommand{\bi}{\begin{itemize}}
\newcommand{\ei}{\end{itemize}}
\newcommand{\bea}{\begin{eqnarray}}
\newcommand{\eea}{\end{eqnarray}}
\newcommand{\ba}{\begin{array}}
\newcommand{\ea}{\end{array}}

\def\ym2{{YM$_2$}}

\def\alphadot{{\dot{\alpha}}}
\def\betadot{{\dot{\beta}}}

\newcommand{\nn}{\nonumber}
\newlength{\sswidth}

\begin{document}
\begin{titlepage}
\begin{flushright}
KIAS-P18006
\end{flushright}
\vskip 1cm
\begin{center}
{\Large \textbf{The topological structure of supergravity:}} 
{\Large \textbf{an application to supersymmetric localization}}
\end{center}
\vspace{1cm}


\begin{center}
  \textsc{  Camillo Imbimbo$^{1,2,a}$, ~Dario Rosa$^{3,b}$}  
\end{center}

\vspace{0.5cm}

\begin{center}

\sl
$^1\,$ Dipartimento di Fisica, Universit\`a di Genova,
Via Dodecaneso 33, Genoa 16146, \rm ITALY \\
\sl $^2\,$ INFN, Sezione di Genova, Via Dodecaneso 33, Genoa 16146, \rm ITALY\\
 \vspace{0.3 cm}
\sl
 $^3$  School of Physics, Korea Institute for Advanced Study (KIAS), \\Seoul 02455, \rm KOREA\\
\vspace{0.3 cm}

\vspace{0.3 cm}
\end{center}

\begin{center}
{\small $^a$camillo.imbimbo@ge.infn.it,  ~~$^b$Dario85@kias.re.kr}
\end{center}

\vspace{1cm}

\centerline{\textsc{ Abstract}}
 \vspace{0.2cm}
 
{\small  The BRST algebra of supergravity is characterized by two different  bilinears of the commuting supersymmetry ghosts: a vector $\gamma^\mu$ and
a scalar $\phi$, the latter valued in the Yang-Mills  Lie algebra. We observe that under BRST transformations $\gamma$ and $\phi$ transform as the superghosts of, respectively, topological gravity and topological Yang-Mills coupled to topological gravity. This topological structure
sitting inside any supergravity leads to universal equivariant cohomological equations for the curvatures  2-forms which hold on supersymmetric  bosonic backgrounds. Additional equivariant cohomological equations can be derived for supersymmetric backgrounds of supergravities  for which certain gauge invariant scalar bilinears of the commuting ghosts exist. Among those, $N= (2,2)$ in $d=2$, which we discuss in detail in this paper, and $N=2$ in $d=4$.
}
\vspace{0.2cm}
\thispagestyle{empty}

\vfill
\eject
\end{titlepage}
\hypersetup{pageanchor=true}
\baselineskip18pt
\tableofcontents
\section{Introduction}

Localization has proven to be  a powerful tool  to study supersymmetric quantum field theories (SQFT)  on curved backgrounds.\footnote{The literature on supersymmetric localization is enormous. See \cite{Pestun:2016zxk} for an extensive overview of the main recent results.}  
To identify localizable backgrounds  one couples supersymmetric matter field theories to classical supergravity: setting
the supersymmetry variations of the fermionic supergravity fields  --- both gravitinos and gauginos ---  to zero, one obtains equations for
the local supersymmetry spinorial parameters, a strategy first exploited in \cite{Festuccia:2011ws}. These differential equations, that are often named {\it generalized Killing spinor equations}, admit non-trivial solutions only for special  configurations
of the bosonic fields of the  supergravity multiplet.  
The relevant supergravity and the particular generalized Killing spinor equations  depend on the global symmetries of the
specific SQFT one is interested in: indeed, conserved currents of the SQFT that one would like to probe couple to gauge fields which must sit
in supergravity multiplets.

It was observed in  \cite{Imbimbo:2014pla} and \cite{Bae:2015eoa} that the generalized Killing spinor equations for certain extended supergravities in 3 and 2 dimensions are equivalent to cohomological equations which are obtained by setting to zero the BRST variations
of fermionic fields of topological gravity coupled to certain topological gauge systems.  In this paper we will provide
a conceptual explanation of the equivalence between the generalized Killing spinor equations of supergravity
and the cohomological equations derived from topological gravity.

Our starting point will be the BRST formulation of supergravity. We will revisit this in Section \ref{sec:sugrabrst}. The BRST structure of a given supergravity theory is specified by the number of its local supersymmetries and by its  bosonic local gauge symmetries. For each local symmetry one introduces ghost fields --- anti-commuting for bosonic local symmetries and commuting for fermionic ones. The bosonic local gauge symmetries always include at least local reparametrizations and local Lorentz transformations: on top of those one can consider additional Yang-Mills local symmetries. These may be  associated, for example,  to {\it global}  R-symmetries of the matter SQFT to which  supergravity can be coupled.  

We will show that the BRST algebra of any supergravity theory takes the form 
\bea
S^2 = \mathcal{L}_\gamma + \delta_{i_\gamma(A) +\phi}
\label{intro:sugraBRSTalgebra}
\eea
Here $S$ is obtained from the nilpotent BRST operator $s$ by subtracting the transformations associated to the bosonic gauge symmetries; $\mathcal{L}_\gamma$
is the Lie derivative along the vector field $\gamma^\mu$  of ghost number 2
\bea
\gamma^\mu = -\frac 1 2  \sum_i \bar\zeta^i\, \Gamma^\mu\, \zeta^i
\label{intro:vectorbilinear}
\eea
where $\zeta^i$, with $i=1,\ldots ,  N$, are the commuting supersymmetry Majorana ghosts\footnote{For concreteness, we are taking  the commuting ghosts to be Majorana spinors in Minkowski space-time here and in the rest of the paper, with the exception of  Section \ref{sec:d2sugra}. We take the Minkowski signature to be $(+, -,\ldots ,-)$.  When $N$ is even it might be more convenient to work with Dirac spinors.   When the space-time dimension permits, one could adapt the discussion to Euclidean signature: this is what we do in Section \ref{sec:d2sugra} where we describe the application to localization of $N = (2,2)$ supergravity in two dimensions. } and $\Gamma^\mu$, with $\mu=1, \ldots , d$,  are the Dirac matrices in curved space-time of dimension $d$; $\delta_{c}$ denotes a gauge transformation with parameter $c$ and $i_\gamma(A)$ is the contraction of the YM gauge field $A$, belonging in the supergravity multiplet, with the vector field $\gamma^\mu$; the scalar field $\phi$  lives in the adjoint representation of the bosonic YM gauge symmetry group.  $\phi$, like $\gamma^\mu$, is a combination of ghost number 2 of the supergravity fields,  {\it bilinear} in the commuting ghosts  $\zeta^i$ and it is independent of the other ghost fields.

The vector ghost bilinear $\gamma^\mu$ and the adjoint-valued scalar ghost bilinear $\phi$ completely specify the supergravity BRST  algebra ---  and thus the supergravity model. 
It should be remarked that the vector ghost bilinear $\gamma^\mu$ has, for any supergravity, the  universal form  (\ref{intro:vectorbilinear}); on the other hand the dependence of the  adjoint-valued scalar ghost bilinear $\phi$  on the the  bosonic fields  is model dependent. We will provide explicit expressions for $\phi$ for the  supergravity theories that
we discuss in this paper.

The algebra (\ref{intro:sugraBRSTalgebra}) encodes the connection between supergravity and topological theories: indeed  both the field  $\gamma^\mu$ and the field $\phi$ admit
a natural topological interpretation on which we will elaborate in Section \ref{sec:topstructure}.
We will show  that, under supergravity BRST transformations, the ghost number 2 vector $\gamma^\mu$ transforms precisely as the superghost of topological gravity while $\phi$ transforms as the superghost of a topological Yang-Mills multiplet. Explicitly, the supergravity BRST  operator $S$  acts on $\phi$  as follows
\bea
&& S \, \phi = i_\gamma(\lambda)\nn\\
&& S\, \lambda = i_\gamma(F) - D\, \phi\nn\\
&& S\, F = - D\, \lambda
\label{intro:topym}
\eea
where $F$ is the field strength 2-form associated to bosonic Yang-Mills symmetries and
$\lambda$ is a ghost number 1 combination of supergravity fields which we will call the (composite) ``topological
gaugino''. $\lambda$ is defined by the supergravity BRST  variation of the gauge field 
\bea
S\, A = \lambda
\eea
The BRST transformations (\ref{intro:topym}) are identical to the  BRST  rules of the topological Yang-Mills multiplet $(F, \lambda, \phi)$ coupled
to topological gravity, once one identifies the supergravity bilinear $\gamma^\mu$ with the topological gravity superghost. The topological gravity
BRST transformations read indeed
\bea
&& S\, g_{\mu\nu} = \psi_{\mu\nu}\nn\\
&& S\, \psi_{\mu\nu} = \mathcal{L}_\gamma\, g_{\mu\nu}\nn\\
&& S\, \gamma^\mu =0
\label{intro:topgrav}
\eea
where $g_{\mu\nu}$ is the metric and $\psi_{\mu\nu}$ is the topological gravitino. The consistency  of (\ref{intro:topgrav}) with (\ref{intro:topym})  hinges on the supergravity transformations for the  supersymmetry ghosts and vierbein 
\bea
&& S\, \zeta^i = i_\gamma(\psi^i) \nn \\
&& S \, e^a =  \sum_i \bar\zeta^i\, \Gamma^a \psi^i
\label{intro:BRSTsugrauniversal}
\eea
where $\psi^i$ are the gravitinos  and $e^a= e^a_\mu\, dx^\mu$ the vierbein 1-forms. Note that the supergravity BRST rules (\ref{intro:BRSTsugrauniversal})  are universal, i.e. independent of the specific supergravity theory.
The fact that the supergravity transformations (\ref{intro:BRSTsugrauniversal}) imply, in particular,  the invariance of the bilinear (\ref{intro:vectorbilinear}) was
observed first in \cite{Baulieu:1985md}.

In Section \ref{sec:topYMtopgr} we will review the coupling, in generic dimensions, of topological gravity to topological YM in any dimensions. This was worked out  in \cite{Imbimbo:2009dy} and \cite{Imbimbo:2014pla} in a three-dimensional context. It is well-known that  the BRST operator of  ``rigid'' topological YM is geometrically the de Rham differential on the
space $\mathcal{A}$ of gauge connections, {\it equivariant} with respect to gauge transformations. Analogously, the BRST operator of topological gravity in (\ref{intro:topgrav}) is the de Rham differential on the space $\mathbf{Met}$ of metrics, equivariant with respect to space-time diffeomorphisms. 
The BRST operator $S$ in  (\ref{intro:topym}), which will be discussed in Section  \ref{sec:topYMtopgr}, is instead the de Rham differential on the space $\mathcal{A}\times\mathbf{Met}$, equivariant with respect to both diffeomorphisms and gauge transformations.  
Independently of the application to supergravity that we consider in this work, this operator should have applications to the study of metric dependence
of Donaldson invariants.

In the rest of this paper we apply to supersymmetric localization the topological structure that we have discovered sitting inside the supergravity BRST algebra. Supersymmetric bosonic backgrounds are obtained by setting to zero the supergravity BRST variations of all the fermionic supergravity fields.  We will refer to the set of such backgrounds as the {\it localization locus}. On the localization locus also the supergravity BRST variations of the {\it composite} fields $\lambda = S \,A$ and $\psi_{\mu\nu}= S\, g_{\mu\nu}$,  must vanish. The resulting equations will be analyzed in Section \ref{sec:universaleqs}. By imposing that the supergravity BRST variation of the topological gravitino vanishes one gets the equation
\bea
&& S \, \psi_{\mu\nu} = \mathcal{L}_\gamma \, g_{\mu\nu} = D_\mu \, \gamma_\nu + D_\nu\, \gamma_\mu =0
\label{intro:Sgravitino}
\eea
which expresses the request that the vector bilinear $\gamma^\mu$ be an isometry of the space-time metric $g_{\mu\nu}$. The fact that, on the localization locus, the vector bilinear $\gamma^\mu$ must be a Killing vector of the space-time metric  is well-known in the supergravity literature. The vanishing
of the supergravity BRST variation of the topological gaugino gives instead the equation
\bea
D\,\phi -i_\gamma(F)=0
\label{intro:Sgaugino}
\eea
This equation is universal, in the sense that it is valid, on the localization locus, for any supergravity in any dimensions. It appears to be a novel
equation which has not been yet explored in either supergravity or topological field theory literature. 

Eq. (\ref{intro:Sgaugino}) takes values in the adjoint representation of the total bosonic YM gauge group: it splits therefore into an equation valued in the Lorentz
local algebra and one in the additional YM symmetry algebra. When either of these are non-abelian, the universal localization  equation (\ref{intro:Sgaugino}) is non-linear. In this case, although  equation (\ref{intro:Sgaugino}) has topological roots, it does not directly defines a cohomological problem. To connect Eq. (\ref{intro:Sgaugino}) to cohomology theory, we need to extract its gauge invariant content. To this end, let us define the {\it generalized} Chern classes 
\bea
\mathbb{C}_n = \mathrm{Tr} (F + \phi)^n = \mathrm{Tr}\,F^n + \cdots + \mathrm{Tr}\,\phi^n 
\eea
which are gauge invariant polyforms. When the backgrounds $F$ and $\phi$ satisfy the localization equation (\ref{intro:Sgaugino}) the generalized Chern classes
obey the equation
\bea
\mathcal{D}_\gamma\,\mathbb C_n=  (d - i_\gamma) \, \mathbb C_n =0
\label{intro:U1equivariantclasses}
\eea
The differential $\mathcal{D}_\gamma \equiv d- i_\gamma$ is the coboundary operator defining the de Rham  cohomology of forms
on space-time,  {\it equivariant} with respect to the  action associated to the Killing vector $\gamma^\mu$. We will
refer to this cohomology as the $\gamma$-equivariant cohomology. Eq. (\ref{intro:U1equivariantclasses}) says
therefore that the generalized Chern classes $\mathbb{C}_n$ are the $\gamma$-equivariant extensions of the ordinary Chern classes $c_n=\mathrm{Tr}\,F^n$.

The ordinary Chern classes $c_n$ are integer-valued when $F$ is the curvature of an hermitian connection. We will compute explicitly the equivariant
extensions of these classes  for $N=(2,2)$ supergravity in $d=2$. It will turn out that  the
equivariant extensions defined by this supergravity are integer-valued as well.  We believe this integrality property is a general  feature of supersymmetric bosonic backgrounds, although we have not yet a proof of this. At any rate,  different values of the $\gamma$-equivariant classes (\ref{intro:U1equivariantclasses}) label different branches of the localization locus. On each of these branches moduli spaces of inequivalent localizing backgrounds may exist--- all with the same (integral) values of the equivariant  Chern classes.

Eqs. (\ref{intro:Sgravitino}), (\ref{intro:Sgaugino}) and its consequence (\ref{intro:U1equivariantclasses}) are obtained by putting to zero the  BRST variation of specific (composite) fermions: the  topological gravitino and gauginos, belonging in topological multiplets
whose ghost number 0 components are (respectively) the metric and the YM curvature.  It should be stressed that
these equations do not, in general,  completely characterize the localization locus. 
One can obtain additional, independent,  gauge invariant cohomological equations for bosonic supersymmetric backgrounds by setting to zero the variations of other, independent,  gauge invariant composite fermions. In Section \ref{sec:scalarghostbilinears} we will discuss how to construct gauge invariant composite fermions out of
the (universal) sector of supergravity which does not involve the auxiliary fields.  These fermions sit in topological gauge invariant multiplets whose ghost number 2 scalars
involve only the supergravity ghosts $\zeta^i$. The 2-form components of these multiplets depend on the auxiliary fields. 
They are not, in general,  curvature of gauge fields associated to local symmetries. To determine them one needs the knowledge of the full off-shell supergravity BRST  rules. 

Multiplets of this kind, that we will call {\it gauge invariant topological multiplets},  can be constructed only for specific supergravity theories with suitable YM local gauge
invariance. In Section \ref{sec:scalarghostbilinears} we will find and discuss examples of these multiplets for $d=2$ $N= (2,2)$ supergravity and $d = 4$ $N=2$ supergravity with $SU(2)$ YM gauge symmetry.

The localization equations associated to the gauge invariant topological multiplets also take the form of $\gamma$-equivariant cohomology equations for classes of degree 2. These equations
are very powerful. They do not appear to have the integrality properties of the cohomology equations associated to the curvature topological multiplets.
Their continuous moduli spaces parametrize the inequivalent localizing backgrounds within a given topological branch classified by the equivariant Chern classes in the equation (\ref{intro:U1equivariantclasses}). We will verify this explicitly for $N=(2,2)$ $d=2$ supergravity in Section \ref{sec:d2sugra}, where we will show that the curvature topological multiplet for this theory  can be expressed  on the localization locus as a quadratic function of the gauge invariant topological multiplets. 

We postpone to future work an analysis of the localization equations associated to the gauge invariant topological multiplets of $N=2$ supergravity with $SU(2)$ YM gauge symmetry in four dimensions. A complete study of these equations requires the knowledge of the off-shell BRST transformations for this model. We believe that such a study might lead to the solution of the long standing problem of the classification of localizing backgrounds for $N=2$ $d=4$ supergravity.

\section{The BRST formulation of supergravity}
\label{sec:sugrabrst}

In the BRST framework one introduces  ghost fields of ghost number +1 in correspondence
to each of the local  symmetries. The {\it bosonic} local symmetries of supergravity
are diffeomorphisms and YM gauge symmetries.  We will denote by $\xi^\mu$ the {\it anti-commuting} vector
ghost field associated to diffeomorphisms, and  by $c$ the {\it anti-commuting} ghost associated to the YM gauge symmetry which takes values
in the adjoint representation of the YM algebra. The YM gauge symmetries always include local Lorentz transformations.
Beyond  local Lorentz gauge symmetry, we will also allow for additional YM gauge symmetries. For the application to localization,
for example, these additional YM gauge symmetries include the R-symmetries of the supersymmetric quantum field theory
whose coupling to supergravity one is considering.  

In correspondence with the $N$ local supersymmetries, one introduces   {\it commuting} supergravity spinorial Majorana ghosts $\zeta^i$, with $i=1,\ldots N$, whose    BRST transformation rules   have the form
\bea
&&s\, \zeta^i = i_\gamma(\psi^i) + \mathrm{diffeos}+ \mathrm{gauge\; transfs}
\label{szeta}
\eea
In this equation  $s$  is the {\it nilpotent} BRST operator
\bea
&& s^2=0,
\eea
$\psi^i= \psi^i_\mu\, dx^\mu$ are the Majorana gravitinos 
and $\gamma^\mu$ is the following vector bilinear of the commuting ghosts\footnote{To avoid confusions, we will denote with $\gamma^\mu$ the ghost bilinear and with $\Gamma^a$ the Dirac matrices.} 
\bea
&&  \gamma^\mu \equiv  - \frac 1 2 \sum_i \bar\zeta^i \,\Gamma^a\, \zeta^i\, e_a^\mu 
\label{gammazetazeta}
\eea
where $e_a^\mu$ are the inverse of the vierbein $e^a\equiv e^a_\mu\, dx^\mu$.  The vector $\gamma^\mu$ has 
ghost number +2. 

Both the Majorana ghosts  $\zeta^i$ and gravitinos $\psi^i$ carry a label $i=1,\ldots N$
on which the  $O(N)$ subgroup of the R-symmetry group acts.  The full R-symmetry group
can however be as large as $U(N)$. 

The  BRST transformations of the
vierbein are universal
\bea
s\, e^a = \sum_i \bar\psi^i \, \Gamma^a \, \zeta^i+ \mathrm{diffeos}+ \mathrm{gauge\; transfs}
\label{svier}
\eea
We will denote the action of  diffeomorphisms with $\mathcal L_\xi$,  the Lie derivative associated with the vector field $\xi^\mu$. The BRST transformations of the diffeomorphism ghost are
\bea
&& s \,\xi^{\mu} = - \frac 12 \mathcal L_\xi \xi^\mu - \frac 1 2 \sum_i \bar\zeta^i \,\Gamma^a\, \zeta^i\, e_a^\mu  
\label{sxi}
\eea
It was noted in \cite{Baulieu:1985md} that the BRST transformations  (\ref{szeta}) and (\ref{svier}) imply that
the  vector ghost bilinear $\gamma^\mu$ transforms as follows
\bea
s\, \gamma^\mu = - \mathcal{L}_\xi\, \gamma^\mu
\label{sgamma}
\eea
This transformation law coincides precisely with the BRST transformation rule for the superghost of topological gravity \cite{Becchi:1995ik}.  Indeed, the full BRST transformations of topological gravity write
\bea
\label{eq:topgravBRST}
&& s \, \gamma^\mu = - \mathcal L_\xi \gamma^\mu \nn\\
&& s \,g_{\mu\nu} = - \mathcal L_\xi g_{\mu\nu} + \psi_{\mu\nu} \ , \nn \\
&& s \,\xi^{\mu} = - \frac 12 \mathcal L_\xi \xi^\mu + \gamma^\mu \nn\\
 && s \, \psi_{\mu\nu} = - \mathcal L_\xi \psi_{\mu\nu} + \mathcal L_\gamma g_{\mu\nu} 
\eea
where $g_{\mu\nu}$ is the metric, $\psi_{\mu\nu}$ is the topological gravitino and $\gamma^\mu$ the vector superghost. From this it is apparent that also the supergravity BRST transformation rules  (\ref{sxi})  match  the
topological gravity ones, once the topological gravity superghost is identified with  the supergravity ghost bilinear  
$\gamma^\mu$ according to (\ref{gammazetazeta}).  The formal coincidence of the topological relations (\ref{eq:topgravBRST}) with the supergravity ones (\ref{sxi}), (\ref{sgamma})  is the   first hint that a topological sector hides behind the supergravity BRST rules. In the following we will bear this to light.

\section{Topological Yang-Mills coupled to topological gravity}
\label{sec:topYMtopgr}

We have seen in the previous Section that  some of the supergravity BRST transformation rules take a form which is identical to the BRST transformations of topological gravity.  In order to uncover the full topological content of the supergravity BRST rules it is necessary  to discuss the coupling of topological gravity to topological Yang-Mills gauge theories. This was  derived, in the form we present here,  in \cite{Imbimbo:2009dy} and \cite{Imbimbo:2014pla} in the specific context of 3-dimensional gauge theories.\footnote{BRST transformations  for topological gravity in $d=4$ coupled to abelian gauge theory were written also in \cite{Baulieu:2002gw}. Although those BRST rules and the one we are presenting in this Section are related by a field redefinition, we note that this field redefinition is essential to obtain {\it gauge covariant} BRST rules, and, thus, gauge covariant localization equations.}  In this  Section we will describe this  construction in generic dimension and explain its geometric meaning.\footnote{A generalization of this construction in $d=2$ has been considered in \cite{Rosa:2016uab}, where the coupling of topological gravity to $2$-dimensional Poisson sigma  models  \cite{Schaller:1994es},\cite{Ikeda:1993fh} has been worked out. } 

The coupling of topological gravity to topological Yang-Mills can be useful in different contexts. The first
one is when both topological gravity and Yang-Mills fields are  dynamical. This gravitational theory  is relevant to study the cohomology of the space of space-time
metrics.  One could also consider the situation in which only the YM degrees of freedom are dynamical while the topological gravity fields play the role of classical backgrounds. In this case the coupling to topological gravity 
probe the dependence of the quantum physical correlators of the YM theory on the background metric. It might be a useful tool to  study, among other things,  the occurence of  possibile quantum  anomalies of the classical topological invariance \cite{Imbimbo:2009dy} or
wall-crossing phenomena in Donaldson theory \cite{Donaldson:1987qho}, \cite{Donaldson:1990kn}, \cite{Moore:1997pc}.  

In the following Sections   topological YM and topological gravity fields will appear as  {\it composites} of the ``microscopic'' supergravity fields. In Section \ref{sec:topstructure} we will explain how such composites emerge from the   BRST formulation of supergravity: we will use the existence of this  topological structure  inside supergravity to  derive general information regarding localization of supersymmetric gauge theories coupled to classical supergravity.
In this Section  --- and only in this Section ---  we think, instead,   of topological YM and topological gravity as fundamental ``microscopic'' theories.

The fields of the topological Yang-Mills theory include, beyond the gauge connection $A=  A^a_\mu\, T^a\,dx^\mu$, the topological
gaugino $\lambda =\lambda^a_\mu\, T^a\, dx^\mu$ of ghost number +1, the gauge ghost $c = c^a \, T^a$ of ghost number $+1$ and the super-ghost $\phi=\phi^a\,T^a$ of ghost number +2.\footnote{The matrices $T^a$, with the index $a$  running over  the adjoint representation of the gauge group, are usually taken to be the generators of the Lie algebra of the gauge group in the fundamental representation.}

The BRST rules of ``rigid'' topological YM ---  i.e. topological YM {\it before} coupling  it to topological gravity --- read
\bea
&& s\, c =-  \frac 1 2 [c, c] + \phi  \nn\\
&& s\, A = -D \, c + \lambda \nn\\
&& s\, \lambda = - D\, \phi - [c, \, \lambda]\nn\\
&& s\, \phi = - [c, \, \phi]
\label{sgaugerigid}
\eea
It is convenient to introduce an operator $S$, whose action is defined on all the fields but not on the ghost field $c$. 
$S$ is related to the nilpotent $s$  by the relation
\bea
S \equiv  s + \delta_c
\label{Sgaugedefinition}
\eea
where $\delta_c$ denotes the gauge transformation with parameter $c$. We will refer to $S$ as the BRST operator {\it equivariant} with respect to local gauge transformations.  The nilpotency of $s$ is equivalent to the algebra
\bea
S^2 =  \delta_\phi
\eea 
It is well known that the operator $S$ defined in (\ref{sgaugerigid}) and (\ref{Sgaugedefinition})  has to be interpreted
as the de Rham differential on the space $\mathcal{A}$ of gauge connections,  equivariant with respect to local gauge transformations. Hence, the BRST cohomology of $S$ corresponds,  geometrically, to the cohomology of $\mathcal{A}$ modulo
local gauge transformations.

Coupling topological YM to topological gravity means to find an extension of the nilpotent BRST operator $s$ which includes local diffeomorphism transformations, with  a ghost parameter $\xi^\mu$ that  transforms according to the BRST rules of topological gravity (\ref{eq:topgravBRST}). 
The request of nilpotency of $s$ dictates the following deformation of (\ref{sgaugerigid})
\bea
&& S\, A = \lambda \nn\\
&& S\, \lambda = i_\gamma \, (F) - D\, \phi\nn\\
&& S\, \phi = i_\gamma(\lambda)
\label{Stotal}
\eea
where $F= d A + A^2$ is the gauge field strength 2-form, $i_\gamma$ denotes the contraction of a form with the vector field $\gamma^\mu$ and $S$ is equivariant both with respect to local gauge transformations and to diffeomorphisms
\bea
S = s + \delta_c + \mathcal L_\xi
\label{eq:Stotaldefinition}
\eea
The new equivariant $S$ of topological YM coupled to topological gravity satisfies the relation
\bea
S^2 =  \delta_{\phi + i_\gamma (A)} + \mathcal{L}_\gamma
\eea
The BRST transformations of the gauge ghost $c$ are
\bea
\label{eq:BRSTtotalghost}
s \, c = - \frac 12 [c, c] - \mathcal{L}_\xi \, c + \phi + i_\gamma (A)
\eea

 The equivariant $S$ defined in  (\ref{Stotal})-(\ref{eq:Stotaldefinition}) is the de Rham differential on the 
product space  $\mathcal{A} \times \mathrm{\bf Met}$, where $\mathrm{\bf Met}$ is the space of space-time metrics,
equivariant with respect to the action of both local gauge transformations and diffeomorphisms. 

We can re-cast the first equation of  (\ref{Stotal}) and the BRST variation of the gauge ghost (\ref{eq:BRSTtotalghost}) in terms of a curvature  $\mathbb{F}$ of the superconnection $c+A$
\bea
\delta \, (c +A) +(c+A)^2 = F + \lambda +\phi \equiv \mathbb{F}
\label{Sgaugecurvature}
\eea 
where
\bea
\delta = s +d+  \mathcal{L}_\xi- i_\gamma
\eea
is nilpotent
\bea
\delta^2 =0
\eea
 The rest of the transformations (\ref{Stotal}) are equivalent to the Bianchi identity
\bea
\delta \, \mathbb{F} + [ c+ A,  \mathbb{F}]=0
\label{StotalconsistencyBianchi}
\eea
Therefore the  gauge invariant  generalized polynomials
\bea
c_n( \mathbb{F}) = \mathrm{Tr}\,  \mathbb{F}^n
\label{genchern}
\eea
are $\delta$-invariant 
\bea
(S + d - i_\gamma)\, c_n( \mathbb{F}) =0
\label{Stotalcninvariance}
\eea
In the context  in which both YM and gravity are dynamical,   Eq.  (\ref{Stotalcninvariance}) expresses
the fact that the generalized forms $c_n ( \mathbb{F})$ encode observables of the theory constructed purely
in terms of  YM degrees of freedom.  In the context in which only the YM degrees of freedom
are dynamical, Eq.  (\ref{Stotalcninvariance})  gives rise to a (classical) Ward identity 
which controls the dependence of quantum correlators of the observables associated to $c_n ( \mathbb{F})$  
on the background metric. In the supergravity context  Eq.  (\ref{Stotalcninvariance})  implies --- as we will show in the following Sections ---  cohomological differential equations  which are satisfied by localizable backgrounds of classical supergravity.

 \section{The topological structure of supergravity}
 \label{sec:topstructure}

The BRST formulation of supergravity involves both  fields with zero ghost number and ghost fields. 
As recalled in Section \ref{sec:sugrabrst},  one must introduce both anti-commuting ghost fields --- which we denoted with  $\xi$ and $c$ --- associated with bosonic 
gauge invariance and commuting ghost fields $\zeta^i$ associated with local supersymmetry.  In the rest of this Section
we will refer, somewhat unorthodoxly,  both to  zero ghost number fields and to commuting ghosts $\zeta^i$ as  ``matter fields''.  We will denote them collectively by $M$.

When acting on matter fields $M$   the nilpotent BRST operator $s$ has the form
\bea
s\, M = - \mathcal{L}_\xi \,M - \delta_c\, M+ \hat{M}(M)
\label{sSM}
\eea
Here $\delta_c\, M$ denotes the gauge group action  on $M$ with parameter $c$. Eq. (\ref{sSM}) defines
therefore the {\it composite} field $\hat{M}(M)$, which is in general a  function of the  {\it matter } fields of the theory --- but not of the anti-commuting ghosts $\xi$ and $c$.
 For example, from the BRST rules (\ref{szeta}), we deduce that
\bea
\hat{\zeta} = i_\gamma(\psi)
\eea
The BRST transformation rules of the bosonic ghost associated to the gauge symmetries have a slightly different structure
\bea
&& s\, \xi^\mu = - \frac 1 2 \mathcal{L}_\xi \, \xi^\mu  + \gamma^\mu\nn\\
&& s\, c = - \frac 1 2 \, [c,c] - \mathcal{L}_\xi \, c + \hat{c}
\label{sghostxichat}
\eea
Here $\gamma^\mu\equiv \hat{\xi}^\mu$ is the universal bilinear defined in (\ref{gammazetazeta}).  The second 
equation defines instead $\hat{c}$ which is a function of ghost number 2 of the matter fields: its  specific
form characterizes the particular supergravity we are considering. 
By imposing nilpotency of $s$ on $M$ 
\bea
&& 0 = s^2\, M
\eea
and using the Jacobi identities associated to gauge and diffeomorphisms transformations
\bea
&&   \frac 1 2  \mathcal{L}_{ \mathcal{L}_\xi \, \xi} \,\phi -  \mathcal{L}_\xi^2 \, \phi =0\nn\\
&& \frac 1 2 [[c,c], \phi] - [c, [c, \phi]]=0
\label{jacobione}
\eea
one obtains the BRST rules for the composite field $\hat{M}(M)$:
\bea
 s\, \hat{M}= -  \mathcal{L}_\xi \,\hat{M} -\delta_c\,  \hat{M}+ \mathcal{L}_\gamma \,M+ \delta_{\hat{c}}\, M
\label{shatM}
\eea
The structure of (\ref{sSM}) and (\ref{shatM}) makes it  convenient to define an operator $S$ \cite{Imbimbo:2009dy}, whose action  ---  defined {\it  on the 
 matter fields only}  --- is obtained
by substracting from $s$ both  diffeomorphisms and gauge transformations
\bea
&& S\, M \equiv  s\, M +  \mathcal{L}_\xi \,M + \delta_c\, M 
\eea
Therefore, by definition, 
\bea
S\, M = \hat{M}(M)
\eea
From (\ref{shatM}) it follows that  the function of the matter fields
\bea
\label{SMhat}
&& S\, \hat{M} = \frac{\partial \hat{M}}{\partial M}(M)\, \hat{M}(M) 
\eea
must satisfy the relation  
\bea
\label{Salgebraone}
S\, \hat{M}= S^2\, M=  \mathcal{L}_\gamma \,M+ \delta_{\hat{c}}\, M
\eea
Therefore nilpotency of $s$ on $M$ implies that the operator $S$ obeys the algebra
\bea
S^2 = \mathcal{L}_\gamma  + \delta_{\hat{c}}
\label{Salgebraonebis}
\eea
Let us remark that (\ref{SMhat}) and (\ref{Salgebraone}) represent  functional differential equations  for
the composites $\hat{M}(M)$.
When imposing nilpotency of $s$ on $\hat{M}$  we obtain instead
\bea
&& S^2\, \hat{M}=  \mathcal{L}_\gamma \,\hat{M}+ \delta_{\hat{c}}\, \hat M+ \mathcal{L}_{S\, \gamma} \, M +   \delta_{S\,\hat{c}}\, M
\eea
We must therefore require 
\bea
&& S\, \gamma^\mu=0 \nn\\
&& S\, \hat{c}=0
\label{SSconsistency}
\eea
The BRST equations (\ref{SSconsistency}) and (\ref{Salgebraone}) constitute a set of functional equations for the all 
the composites  ---  $\gamma^\mu$, $\hat{c}$ and $\hat{M}$.  Constructing a supergravity theory amounts, in essence,
in solving such functional equations. 

We observed already in the previous Section that the universal composite defined  in (\ref{gammazetazeta}) does indeed
satisfy  the first of Eqs.  (\ref{SSconsistency}). Let us work out the constraints on the form of $\hat{c}$ which follow from the second equation in (\ref{SSconsistency}).
The nilpotency of $s$ on the gauge symmetry ghost $c$ gives 
\bea
&& 0 =s^2 \, c =  [c,- \frac 1 2 \, [c,c] - \mathcal{L}_\xi \, c + \hat{c} ] - \mathcal{L}_{  - \frac 1 2 \mathcal{L}_\xi \, \xi^\mu  + \gamma^\mu} \, c  +\nn\\
&&\qquad +\mathcal{L}_\xi \,( - \frac 1 2 \, [c,c] - \mathcal{L}_\xi \, c + \hat{c}) + s\, \hat{c}\nn\\
&&\qquad = -[c,  \mathcal{L}_\xi \, c] +[c,  \hat{c} ]+  \frac 1 2 \mathcal{L}_{    \mathcal{L}_\xi \, \xi^\mu}\, c  - \mathcal{L}_\gamma  \, c  +\nn\\
&&\qquad -[ \mathcal{L}_\xi  \, c,c] - \mathcal{L}_\xi^2 \, c + \mathcal{L}_\xi\, \hat{c} + s\, \hat{c}\nn\\
&&\qquad=  [c,  \hat{c} ] - \mathcal{L}_\gamma  \, c  + \mathcal{L}_\xi\, \hat{c} + s\, \hat{c}
\eea
where we again used (\ref{jacobione}).  Hence we obtain
\bea
s\, \hat{c} = -   \mathcal{L}_\xi\, \hat{c} -[c,  \hat{c} ] + \mathcal{L}_\gamma  \, c 
\label{shatcone}
\eea
On the other hand,  since $\hat{c}$ is a composite satisfying the second equation in (\ref{SSconsistency}), it must be that
\bea
0= S\, \hat{c} = s\, \hat{c} +   \mathcal{L}_\xi\, \hat{c} + \delta_c\, \hat{c}
\label{Shatctwo}
\eea
Comparing (\ref{shatcone}) with (\ref{Shatctwo})  one deduces  the transformation rules of the functional $\hat{c}$ under bosonic gauge symmetry  
\bea
\delta_c \hat{c} =  -\mathcal{L}_\gamma  \, c + [c, \hat{c}]
\eea
In other words,  $\hat{c}$ does {\it not} transform homogeneously under gauge transformations. It must then have the general form
\bea
\hat{c} = i_\gamma(A) + \phi
\eea
where $A$ is the gauge field 1-form associated to the algebra of the  bosonic YM gauge invariances (which include the local Lorentz transformations) and $\phi$ is a composite
fields with values in the adjoint of the YM Lie algebra. $\phi$  transforms homogeneously under gauge transformations.    

Summarizing, the algebra satisfied by $S$ is
\bea
&& S^2 = \mathcal{L}_\gamma + \delta_{ i_\gamma(A)+ \phi}
\label{Salgebra}
\eea
The consistency condition $S \, \hat{c}=0$ translates into the equation
\bea
 S\, \phi=  i_\gamma(S\,A) 
\eea
The composite $ S\, A = \hat{A}$ is  the topological gaugino, which will be denoted by $\lambda$: 
\bea
S\, A = \lambda
\eea
$A$ and $\lambda$ sit into a multiplet with values in the adjoint of the gauge algebra, 
\bea
&& S\, A = \lambda \nn\\
&& S\, \lambda = i_\gamma \, (F) - D\, \phi\nn\\
&& S\, \phi = i_\gamma(\lambda)
\label{Sgauge}
\eea
These supergravity BRST transformation rules coincide with the BRST rules of topological YM coupled to topological gravity that we wrote in (\ref{Stotal}). 
Both the topological gaugino $\lambda$ of ghost number +1 and the topological Yang-Mills superghost $\phi$ of ghost number +2  are composite
fields in terms of the supergravity fields.  This composite topological  multiplet represents the universal topological sector which sits inside  generic supergravity. 

To give a concrete example, let us consider the $d=4$, $N=1$ ``new minimal'' supergravity.\footnote{In the following paragraphs of this Section we consider the Minkowskian theory and use conventions and notations of \cite{Ferrara:1988qxa}. In particular the normalization of the commuting ghosts and gravitinos differ by those used in the rest of our paper.} The local bosonic YM simmetries
of this theory are local Lorentz transformations and local $U(1)_R$ R-symmetry. Let $c$ and $c^{ab}$ be
the corresponding  anti-commuting  ghosts, with $a,b=1,\ldots 4$. The bosonic local simmetries
act on the commuting Majorana spinorial ghost $\zeta$ as follows
\bea
\delta_c\, \zeta = \bigl(\frac i 2 \, c\, \Gamma_5+ \frac i 4 c^{ab} \, \sigma_{ab}\bigr)\,\zeta
\eea
Let  $\omega^{ab}= dx^\mu \omega_\mu^{ab}$ be the spin-connection and  $A^{(R)}= dx^\mu\,A^{(R)}_\mu$ the $U(1)_R$ gauge field. The 1-form connection with values in the  total bosonic YM Lie algebra is
\bea
A = - \bigl(\frac i 2 \, A^{(R)} \, \Gamma_5+ \frac i 4 \omega^{ab} \, \sigma_{ab}\bigr)
\eea

The BRST transformations of the Majorana gravitino field  $\psi= dx^\mu\, \psi_\mu$ take the form
\bea
S\, \psi= - (d + A^+) \, \zeta = - D^+\, \zeta
\label{N1d4gravitinoone}
\eea
where
\bea
A^+=  A+ e^a\, \bigl(\frac{i}{2}\, \Gamma_5\, H_a - \frac{i}{4}\, \epsilon_{abcd}\, \sigma^{cd}\, H^b\bigr)
\label{N1d4gravitinotwo}
\eea
is a 1-form with values in the total bosonic gauge Lie algebra,  $H_\mu dx^\mu $  is an auxiliary   1-form field\footnote{The auxiliary field $H_a$  is constrained to have zero divergence, up
to fermionic terms:  more precisely its Hodge dual  $\star H$  satisfy: $\star H = d \, B + \frac{i}{8}\, \bar\psi \,\Gamma\, \psi$
where $B$ is a 2-form.}, and
\bea
H_a = e^\mu_{\,a} \, H_\mu
\eea
The  BRST transformations of $H_a$ are\bea 
&& S\, H_a = -\frac{i}{8}\,  \epsilon_{abcd}\,\bar\zeta\, \Gamma^b\, \psi^{cd}
\eea
where $\psi_{ab} \equiv e^\mu_a\, e^\nu_b \, \psi_{\mu\nu} = e^\mu_a\, e^\nu_b \, \bigl(D^+_\mu\, \psi_\nu-D^+_\nu\, \psi_\mu)$.

Recalling  the Fierz identity valid for Majorana spinors in four dimensions
\bea
\bar\zeta\, \Gamma^a\, \zeta \, \Gamma_a\, \zeta =0
\eea
one verifies that $S$ satisfies the algebra (\ref{Salgebra})  with 
\bea
\phi=-i_\gamma(e^a)\, \bigl(\frac{3\,i}{2}\, \Gamma_5\, H_a + \frac{i}{4}\, \epsilon_{abcd}\, \sigma^{cd}\, H^b\bigr)
\eea
Thus,  the  composite superghost $\phi$ for $N=1$ $d=4$ supergravity is $i_\gamma$-trivial
\bea
\phi= -i_\gamma(\Delta) \qquad \Delta \equiv\frac{3\,i}{2}\, \Gamma_5\, H + \frac{i}{4}\, \epsilon_{abcd}\, \sigma^{cd}\, e^a\,H^b
\eea
When $\phi$ is $i_\gamma$-exact, one can introduce a connection
\bea
A^- \equiv A - \Delta
\eea
whose  associated composite gaugino is
\bea
\lambda^-\equiv S\,A^- = \lambda - S\, \Delta
\label{lambdaminus}
\eea
with
\bea
i_\gamma(\lambda^-) = 0
\eea
From the algebra (\ref{Salgebra}) we obtain
\bea
&& S\, \lambda^- = i_\gamma(F) - D\, \phi - \mathcal{L}_\gamma \,\Delta - [i_\gamma(A^-), \Delta]= \nn\\
&&\qquad = i_\gamma(F^-) 
\eea
where
\bea
F^- \equiv d A^- + A^{-\;2}
\eea
is the curvature of the connection $A^-$. 

We see therefore that when $\phi$ is $i_\gamma$-trivial, there exists a connection $A^-$ and a corresponding topological
multiplet $\mathbb{F}^- = F^- + \lambda^- + \phi^-$, with vanishing ghost number 2  component, $\phi^-=0$. This is a special feature of $N=1$ $d=4$ supergravity: we will see in the next Sections that there exist supergravity models, with extended supersymmetry, for which $\phi$ is not $i_\gamma$-exact. 

\section{The universal cohomological equations for supersymmetric backgrounds}
\label{sec:universaleqs}

Bosonic supergravity configurations which are invariant under supersymmetry trasformations define classical backgrounds
to which supersymmmetric quantum field theories (SQFTs)  can be coupled.  SQFT coupled to such backgrounds are localizable, which means that their partition function is one-loop exact.   Supergravity bosonic backgrounds invariant under supersymmetry are identified
by  generalized convariantly constant spinor equations for the supergravity ghosts $\zeta^i$, which are obtained by setting to zero
the supersymmetric variations of the  spinorial fermionic fields --- gravitinos and gauginos of supergravity.

The results of the previous section show that localizable supergravity backgrounds satisfy universal  equations
obtained by setting to zero the BRST variations of the {\it composite} topological gravitinos and gauginos defined in (\ref{eq:topgravBRST}) and (\ref{Sgauge}).  Vanishing of the BRST variation of the composite topological gravitino leads to
\bea
\label{eq:cohomoeqsgrav}
&& D^\mu\, \gamma^\nu + D^\nu\, \gamma^\mu  =0
\eea
This  equation says that a necessary condition for  localization is that  the composite vector ghost bilinear (\ref{gammazetazeta})  be  a  Killing vector of the space-time metric $g^{\mu\nu}$, a condition which is  well-known in the localization literature. 

The vanishing of the BRST variation of the composite topological gauginos leads instead to the equation
\bea
&&  D\, \phi- i_\gamma \, (F)=0 
\label{eq:cohomoeqsgauge}
\eea
where $\gamma^\mu$ is the Killing vector in Eq. (\ref{eq:cohomoeqsgrav}) and $\phi$ is the  bilinear of supergravity ghosts which we defined in the previous Sections and which characterizes the
supergravity BRST algebra (\ref{Salgebra}).  

The topological equations (\ref{eq:cohomoeqsgrav}-\ref{eq:cohomoeqsgauge}) for the generalized Killing spinors    $\zeta^i$  are {\it universal}, in the sense that they take the same form in any dimensions and for any supergravity model, unlike the generalized Killing spinor equations. 

It should be emphasized that  equations (\ref{eq:cohomoeqsgauge}) are obtained
by setting to zero the supergravity BRST variation of a specific (composite) fermion --- the topological gaugino $\lambda$.
These equations therefore do not, in general,  completely characterize the localization locus. There might be more independent
equations valid on the localization locus, obtained by setting to zero the BRST variation of other (composite) fermions. 
For example, for the $N=1$ $d=4$ new minimal supergravity, on top of equation (\ref{eq:cohomoeqsgauge}), associated to the topological multiplet
of the curvature $F$, setting to zero  the BRST variation of the gaugino $\lambda^-$ in (\ref{lambdaminus}), one obtains the equation
\bea
\label{eq:F-horizontal}
&& S \, \lambda^- = i_\gamma(F^-) =0
\eea
valid for supersymmetric backgrounds. 
Coming back to the general case, let us remark that in equation (\ref{eq:cohomoeqsgauge}), $F$ and $\phi$ take values in the adjoint representation of the local bosonic gauge symmetries.  Therefore Eqs. (\ref{eq:cohomoeqsgrav}) split into two separate sets of equations with the same form: one associated to    local Lorentz  symmetry  and the other with  additional local YM  symmetries. 

When either of these symmetries are not abelian Eqs. (\ref{eq:cohomoeqsgrav}) are not gauge invariant: 
Their gauge invariant content is captured by the equations satisfied by the generalized Chern classes (\ref{genchern}):
\bea
(d - i_\gamma)\, c_n( \mathbb{F})=0 
\label{eq:chernequivarianteq}
\eea   
where 
\bea
\label{eq:chernequivariant}
c_n = \mathrm{Tr} \, \bigl[F^n +n\, F^{n-1}\, \phi+ \cdots + \phi^n\bigr]\qquad n=1, 2, \ldots
\eea
and $\gamma^\mu$ is the Killing vector satisfying (\ref{eq:cohomoeqsgrav}).

We see therefore that  the generalized Chern classes evaluated for localizing backgrounds  are closed under
the  coboundary operator
\bea\label{eq:coboundaryDgamma}
\mathcal{D}_\gamma = d - i_\gamma \qquad \mathcal{D}_\gamma^2 = - \mathcal{L}_\gamma
\eea
 associated to the de Rham cohomology  of forms on space-time,  {\it equivariant} with respect to the action  of the Killing vector $\gamma^\mu$.  We will call the cohomology relative to the coboundary operator
 $\mathcal{D}_\gamma$ the $\gamma$-equivariant (polyform) cohomology.
 
 The  $\gamma$-equivariant classes defined by  the $c_n$'s  are {\it invariants} of the localization backgrounds: they are the same
for backgrounds which are equivalent under local BRST transformations of supergravity. In other words  the  classes associated to the $c_n$'s are functions of the moduli of the space of inequivalent localizable backgrounds. 

 On the other hand it is possible that inequivalent localizable backgrounds give rise to $c_n$'s which are different representatives of the 
 same $\gamma$-equivariant class. In the next Section, we will consider  more in\-dependent, gauge invariant, com\-posite fermions which can be defined for certain supergravity models.  Setting to zero their BRST variations one obtains  additional topological equations satisfied by supersymmetric backgrounds. In Section \ref{sec:d2sugra} we will show that  these equations allow for a finer classification of the inequivalent localizable backgrounds.

\section{The gauge invariant ghost bilinears}
\label{sec:scalarghostbilinears}

We have seen in  Section \ref{sec:topstructure} that for any supergravity theory there exists a scalar ghost bilinear $\phi$ of ghost number +2 with values in the Lie algebra of the bosonic gauge symmetries which characterizes the BRST algebra (\ref{Salgebra}).  The ghost bilinear $\phi$ is in general a functional of both the supergravity ghosts $\zeta^i$ and the bosonic  fields of ghost number 0 sitting in the supergravity multiplet.  The key property of $\phi$, which can be read off Eq. (\ref{Sgauge}),  is that its BRST variation is $i_\gamma$-exact:
\bea
S\, \phi = i_\gamma(\lambda)
\label{phiBRST}
\eea
This property ensures that, on the localization locus,  $\phi$ satisfies the topological equation (\ref{eq:cohomoeqsgauge}). 

Let us recall that the BRST variation of the supergravity ghosts $\zeta^i$ is also $i_\gamma$-exact
\bea
S\, \zeta^i = i_\gamma(\psi^i)
\label{Szetabis}
\eea
It follows that scalar and gauge invariant ghost bilinears which are {\it indepedent} of extra bosonic
fields automatically satisfy (\ref{phiBRST}). Hence, they give rise to cohomological equations of the form (\ref{eq:cohomoeqsgauge}).   We will consider in this paper two supergravity models for which ghost bilinears of this sort can be constructed.

 The first one is  $N=(2, \, 2)$ supergravity in $d=2$, where the gauge symmetry group is $SO(2)_R$. In this case it is convenient to collect the two Majorana ghosts $\zeta^i$, with $i=1,2$, into one single  Dirac ghost $\zeta$ on which the gauge group $SO(2)_R\sim U(1)_R$  acts by multiplication by a real phase.\footnote{The following discussion is valid for both Minkowski and Euclidean signature.} Then the two scalar bilinears
 \bea
 \varphi_1 =\bar\zeta \, \zeta\qquad \varphi_2= \bar\zeta \, \Gamma_3\, \zeta
 \label{c0c0tilde}
 \eea
are gauge invariant and thus $S$-invariant modulo $i_\gamma$:
\bea
&& S\, \varphi_i = i_\gamma(\lambda_i)\qquad i=1,2
\eea
where 
\bea
\lambda_1 \equiv \bar\psi \,\zeta+ \bar\zeta\,\psi\qquad \lambda_2 \equiv \bar\psi\,\Gamma_3 \zeta+ \bar\zeta\,\Gamma_3\psi
\eea
It follows from the BRST algebra (\ref{Salgebra}) that  the generalized forms
\bea
\mathbb{H}_i\equiv \phi_i+ \lambda_i + \hat{H}^{(2)}_i
\eea
satisfy
\bea
\delta\, \mathbb{H}_i= (S + d -i_\gamma) \, \mathbb{H}_i=0
\eea
The 2-forms $\hat{H}^{(2)}_i$ write
\bea
\hat{H}^{(2)}_1 = \bar\psi \, \psi + H^{(2)}_1\qquad \hat{H}^{(2)}_2 =\bar\psi \, \Gamma_3\,\psi + H^{(2)}_2
\eea
where $H^{(2)}_i$, with $i=1,2$, are the graphiphoton field strengths. 

As explained in the previous Section, on the localization locus the following cohomological equations hold
\bea
d\, \varphi_i - i_\gamma(H^{(2)}_i)=0
\label{ghostbieqs}
\eea

Scalar ghost bilinears of the same kind can be constructed also for
$N=2$ $d=4$ supergravity in which the R-symmetry  $SU(2)_R$ is gauged. In this case we can take the commuting
supersymmetry ghosts to be two-components spinors $\zeta_\alpha^i$ where $\alpha=1,2$ is the Lorentz spinorial index  and $i=1,2$ is the index of the fundamental representation of the gauge group $SU(2)_R$, together with their conjugate $\bar\zeta^\alphadot_i$.  Two independent scalar
and $SU(2)_R$-invariant ghost bilinears are 
\bea
&& \varphi= \epsilon^{\alpha\beta}\, \epsilon_{ij}\, \zeta^i_\alpha\,\zeta^j_\beta\nn\\
&& \bar{\varphi}=\epsilon_{\alphadot\betadot}\, \epsilon^{ij}\, \bar\zeta_i^\alphadot\,\bar\zeta_j^\betadot
\eea
Again, thanks to  (\ref{Szetabis}), both $\varphi$ and $\bar{\varphi}$ are $S$-invariant modulo $i_\gamma$ 
\bea
&& S\,  \varphi= i_\gamma( \epsilon^{\alpha\beta}\, \epsilon_{ij}\, \psi^i_\alpha\,\zeta^j_\beta+\epsilon^{\alpha\beta}\, \epsilon_{ij}\, \zeta^i_\alpha\,\psi^j_\beta) \equiv i_\gamma(\Lambda)\nn\\
&& S\,   \bar\varphi=i_\gamma(\epsilon_{\alphadot\betadot}\, \epsilon^{ij}\, \bar\psi_i^\alphadot\,\bar\zeta_j^\betadot+\epsilon_{\alphadot\betadot}\, \epsilon^{ij}\, \bar\zeta_i^\alphadot\,\bar\psi_j^\betadot)\equiv i_\gamma(\bar\Lambda)
\eea
where $\psi^i_\alpha$ and $\bar\psi_i^\alphadot$ are the gravitinos. The algebra  (\ref{Salgebra})  ensures that  $ \varphi$ and $ \bar\varphi $ sit in abelian topological gauge multiplets
\bea
&& \mathbb{T}\equiv \varphi+ \Lambda+ T^{(2)}\nn\\
&& \bar{\mathbb{T}}\equiv \bar{\varphi}+ \bar{\Lambda}+ \bar{T}^{(2)}
\eea
satisfying
\bea
 (S + d -i_\gamma) \, \mathbb{T}=0\qquad (S + d -i_\gamma) \, \bar{\mathbb{T}}=0
\eea
Obtaining the 2-forms $ T^{(2)}$ and $\bar{T}^{(2)}$ as functionals of the fields of the supergravity multiplets requires  the knowledge of the off-shell BRST transformations of $N=2$ $d=4$ Poincar\'e supergravity with gauge
group $SU(2)_R$.  Since these do not seem to be readily available in the literature we will present this calculation 
elsewhere. At any rate one can anticipate that  the following cohomological $\gamma$-equivariant equations will
hold on the localization locus of $N=2$, $d=4$ supergravity
\bea
d\, \varphi - i_{\gamma}(T^{(2)})=0\qquad d\, \bar{\varphi} - i_{\gamma}(\bar{T}^{(2)})=0
\label{eq:equivariantvarphiT}
\eea
We expect these equations to play a central role in understanding the space of localizing backgrounds
of $N=2$ $d=4$ Poincar\'e supergravity.

\section{An example: $d=2$ $N=(2,2)$ supergravity }
\label{sec:d2sugra}

 In this Section we will  work out the details of the  topological structure of  $N= (2,2)$  supergravity in $d=2$  
 with Euclidean signature.\footnote{For a description of $N = (2,2)$ supergravity in two dimensions see, for example, \cite{Gates:1988ey}.} The moduli space of localization backgrounds for this  supergravity theory has been fully analysed and described in \cite{Bae:2015eoa}, extending the results of \cite{Closset:2014pda}. In the following we will see how the analysis 
of \cite{Bae:2015eoa}  fits into the framework developed in this paper.

The ghost bilinear $\phi$, characterizing the BRST algebra  (\ref{Salgebra})  of $N=(2,2)$ supergravity in $d=2$, 
has the general form 
\bea
\phi= \phi_{R}\, (- i\, \mathbb{I}) + \phi_{\mathrm{Lor}} \, \bigl(-\frac i 2\,\Gamma_3 \bigr)
\eea
where $-i\, \mathbb{I}$ is the generator of the vectorial $U(1)_R$ gauge transformations on  Dirac spinors, 
 $ -\frac i 2\,\Gamma_3$ is the generator of the local Lorentz transformations, $\phi_R$ is the scalar ghost bilinear associated
 to the $U(1)_R$ gauge symmetry  and  $ \phi_{\mathrm{Lor}}$  the one relative to the Lorentz local transformations.
 The  BRST transformations
of the gravitino Dirac field $\psi= dx^\mu\, \psi_\mu$ write
\bea
\label{eq:BRSTgravitino2d}
&& S \, \psi  
= - D \, \zeta - \frac i2\, H_2 \, dx^\mu \, \Gamma_\mu \zeta  - \frac i2\, H_1 \, dx^\mu \, \Gamma_3 \Gamma_\mu \zeta \ ,
\eea
where 
\bea
\label{eq:Dzetadef2d}
D \, \zeta \equiv dx^\mu \, \bigl( \partial_\mu - \frac i2\,\omega_\mu\,\Gamma_3  - i \, A_\mu \bigr) \, \zeta \ .
\eea
and $H_i= \star H^{(2)}_i$, with $i=1,2$,  are the scalars dual to the field strengths of the two
graviphotons.  Taking into account the Fierz identity
\bea
&&  \Gamma_\mu \,\zeta \bar \zeta \, \Gamma^\mu \zeta =  (\varphi_1 - \varphi_2 \, \Gamma_3) \, \zeta 
\eea
one  derives from   $S^2 \, \zeta$ the values for the ghost bilinears $\phi_R$ and $ \phi_{\mathrm{Lor}}$
\bea
\label{eq:superghostsgauge_2d}
&& \phi_{\mathrm{Lor}} = \mathbb{R}^{(0)} = \eta^{ij}\,\varphi_i\, H_j  \nn\\
 && \phi_R=  \mathbb{F}^{(0)}_R=  \frac 1 2\, \epsilon^{ij}\, \varphi_i\, H_j
 \label{N2d2phis}
\eea
where $\eta^{ij}$ is the Lorentzian metric $\eta^{11}=- \eta^{22}=1$
and $\epsilon^{ij}$ is the Levi Civita tensor in 2 dimensions.
The universal topological equations (\ref{eq:cohomoeqsgauge}) for supersymmetric backgrounds read
\bea
&& d\, \phi_{\mathrm{Lor}} - i_\gamma(R^{(2)})=0\nn\\
&& d\, \phi_R - i_\gamma(\mathcal{F}_R^{(2)})=0
\label{2Nd2:universalghosteqs}
\eea
These equations 
mean that the polyforms of degree 2 associated to the curvature and the field
strength of the $U(1)_R$ gauge field
\bea
&&  \mathbb{R} =  \phi_{\mathrm{Lor}} + R^{(2)}\nn\\
&& \mathbb{F}_R = \phi_R+ \mathcal{F}_R^{(2)}
\eea
 are $\gamma$-equivariantly closed:
 \bea
 \mathcal{ D}_\gamma \, \mathbb{R} = 0\qquad \mathcal{ D}_\gamma \, \mathbb{F} = 0
 \label{2Nd2:universalghosteqsbis}
 \eea
We explained in Section \ref{sec:scalarghostbilinears} that, for $N= (2,2)$ $d=2$ supergravity, one can construct, starting from the gauge invariant ghost bilinears (\ref{c0c0tilde}), two more  equivariant forms $\mathbb {H}_i$  of degree 2, which  satisfy the $\gamma$-equivariant cohomology equations  (\ref{ghostbieqs})
\bea
 \mathcal{ D}_\gamma \, \mathbb{H}_i = 0
 \label{ghostbieqsbis}
 \eea
on the localization locus.

The relation (\ref{N2d2phis}) between the scalar components of the equivariantly closed forms  $\mathbb{H}_i$, 
$\mathbb{R}$ and $\mathbb{F}_R$ can be extended to the  following relations between polyforms
\bea
&&\mathbb{F}_R =  \frac 1 2\, \epsilon^{ij}\, \mathbb{H}_i \,L\,(\mathbb{H}_j)\nn\\
&& \mathbb{R}= \eta^{ij}\, \mathbb{H}_i \, L\,(\mathbb{H}_j) 
\label{N2d2compositeness}
\eea
where the equivariant form $L\,(\mathbb{H}_j)$ is defined in Eq. (\ref{app:Loperator}) of the Appendix \ref{app:eqcohomology2}. These relations show that the universal equations for the curvatures (\ref{2Nd2:universalghosteqsbis}) are in fact consequence
of the equations for the gauge invariant ghost bilinears  (\ref{ghostbieqsbis}). They also connect the cohomology classes of the $\mathbb{H}_i$ with those of
 $\mathbb{R}$ and $\mathbb{F}_R$.   To see this, let us  introduce one more equivariantly closed polyform
 \bea
 && \hat{\mathbb{R}} = - \eta^{ij}\, \Phi_i \,\Phi_j = 
 -\bigl(\gamma^2  + k\, d \gamma^2 \bigr)
 \label{Fierzeqbis}
 \eea
 where we used the Fierz identity $\varphi_1^2- \varphi_2^2= \gamma^2$. Note that, as shown in Appendix \ref{app:eqcohomology2},
 \bea
 \mathbb{R} = L\, (\hat{\mathbb{R}})
\label{PhiLhatPhiLbis}
\eea
It follows from (\ref{Fierzeqbis}) that
 \bea
 \hat{\mathbb{R}}(p_\pm) =\mathbb{H}_1(p_\pm) \,\mathbb{H}_1 (p_\pm)-\mathbb{H}_2(p_\pm) \,\mathbb{H}_2 (p_\pm)=0
\eea
  Hence  $ \hat{\mathbb{R}}$ is cohomologically trivial\footnote{As mentioned in the Appendix \ref{app:eqcohomology2}, $\hat{\mathbb{R}}$  and $\mathbb{R}= L(\hat{\mathbb{R}})$ do not need to be cohomologous.} and 
  \bea
  \big|\mathbb{H}_1(p_\pm)\bigr| = \big|\mathbb{H}_2(p_\pm)\big| 
  \eea
  Let us therefore put
  \bea
  \mathbb{H}_1(p_\pm)= \sigma_\pm\, \mathbb{H}_2(p_\pm)
  \eea
  where $\sigma_\pm$ is the relative sign between $\mathbb{H}_1(p_\pm)$ and $\mathbb{H}_2(p_\pm)$.
  Moreover, from the second of Eqs (\ref{N2d2compositeness}) we  obtain
  \bea
 L\, (\mathbb{H}_1)(p_\pm) = \frac{ \mathbb{R}(p_\pm) }{\mathbb{H}_1(p_\pm)}+ \frac{\mathbb{H}_2(p_\pm)}{\mathbb{H}_1(p_\pm)}\, L\,(\mathbb{H}_2(p_\pm))
\eea
 Plugging this inside the first of Eqs.  (\ref{N2d2compositeness}) one arrives to:\footnote{From this relation we derive in particular that $ \int_{S^2}\, \mathcal{F}^{(2)}_R= 
 \frac 1 2 \,\bigl(\sigma_+ +\sigma_-\bigr)$. Therefore, the first Chern class of the gauge field can take values $-1,0,1$ according to the signs $\sigma_\pm$, in agreement with \cite{Bae:2015eoa} and \cite{Closset:2014pda}.}
  \bea
  &&\mathbb{F}_R(p_\pm)=  \frac 1 2 \,\bigl(\mathbb{H}_1(p_\pm)\, L\,(\mathbb{H}_2)(p_\pm)-\mathbb{H}_2(p_\pm)\, L\,(\mathbb{H}_1)(p_\pm)\bigr)=\nn\\
 && \qquad =-\frac 1 2 \, \frac{\mathbb{H}_2 (p_\pm)}{\mathbb{H}_1(p_\pm)}\, \mathbb{R}(p_\pm)= -\frac 1 2 \,\sigma_\pm\, \mathbb{R}(p_\pm) = \pm \frac 1 2 \,\sigma_\pm
 \label{RFchern}
  \eea
  Eqs. (\ref{RFchern}) and (\ref{Rchern}) show that the $\gamma$-equivariant cohomology classes of both the curvatures  polyforms $\mathbb{F}_R$ and $\mathbb{R}$ are (semi)-integral. This should be contrasted with the cohomology classes
  of the $\mathbb{H}_i$'s which, as shown in \cite{Bae:2015eoa}, depend on a continuous parameter labelling inequivalent localization backgrounds.  We believe that this phenomenon is a general property of 
  the $\gamma$-equivariant curvature  polyforms of supergravity, although we do not have yet a general proof of it.

\section{Conclusions}
\label{sec:conclusions}

In this paper we showed that a topological structure sits inside supergravity. 
We did this by writing the BRST algebra of  supergravity in the form
\bea
S^2 = \mathcal{L}_\gamma + \delta_{i_\gamma (A) + \phi} \ ,
\label{con:brstalgebra}
\eea
where $S$ is the supergravity BRST operator --- equivariant with respect both gauge transformations and diffeomorphisms --- $\mathcal L_\gamma$ is the Lie derivative along the vector $\gamma \equiv \gamma^\mu \, \frac{\partial}{\partial \, x^\mu}$  and $\delta_c$ denotes a gauge transformation with parameter $c$. The two fields $\gamma$ and $\phi$ are {\it bilinears} of the commuting supersymmetry ghosts $\zeta^i$, where $i=1,\ldots N$ and $N$ is the number of local supersymmetries. The vector field $\gamma$ is given by
\bea
\label{eq:concgamma}
\gamma^\mu = -\frac 12 \sum_i \bar \zeta^i \gamma^\mu \zeta^i
\eea
This expression is  {\it universal} in the sense that it is valid  for any supergravity theory in any dimension. On the other hand, the scalar field $\phi$, which is valued in the adjoint representation of the bosonic YM gauge symmetry group, 
 is a ghost bilinear whose dependence on the bosonic fields of supergravity is  {\it non-universal}, i.e. it is  theory dependent.

Our central observation is that both $\gamma^\mu$ and $\phi$ have a topological meaning: they can be identified,
respectively, with the superghost of topological gravity and the superghost of a  topological Yang-Mills theory whose gauge group is the product of the local  Lorentz transformations and the Yang-Mills symmetries of the supergravity theory under consideration.  

This identification relies on the fact that the supergravity BRST transformations of the ghost bilinears $\gamma^\mu$ and $\phi$
coincide precisely with the BRST rules of topological gravity 
\bea
&& S\, g_{\mu\nu} = \psi_{\mu\nu}\nn\\
&& S\, \psi_{\mu\nu} = \mathcal{L}_\gamma\, g_{\mu\nu}\nn\\
&& S\, \gamma^\mu =0
\label{eq:conctopgrav}
\eea
and of topological Yang-Mills coupled to topological gravity:
\bea
&& S\, A = \lambda \nn\\
&& S\, \lambda = i_\gamma \, (F) - D\, \phi\nn\\
&& S\, \phi = i_\gamma(\lambda)
\label{eq:concSgauge}
\eea

The BRST variations (\ref{eq:concSgauge}) are not familiar in the topological Yang-Mills literature, although they already appeared in \cite{Imbimbo:2014pla} in a three-dimensional context. In this paper we wrote them  down for any dimensions and discussed their geometrical interpretation. We believe that these transformations could be of interest to study, for example, the metric dependence of Donaldson invariants --- regardless of the application to
supergravity that we explored in this article.

It was observed in \cite{Imbimbo:2014pla} and \cite{Bae:2015eoa} that the conditions for unbroken supersymmetry in certain off-shell supergravities with extended supersymmetry can be recast in terms of cohomological equations. In \cite{Imbimbo:2014pla} and \cite{Bae:2015eoa} these cohomological equations were obtained   by setting to zero the BRST variations of fermionic fields of topological gravity coupled to some additional topological gauge multiplets. 

In this paper we gave a conceptual explanation of this equivalence, by exploiting the topological structure
of supergravity captured by the BRST transformations laws (\ref{eq:conctopgrav}) and (\ref{eq:concSgauge}).
By setting to zero the BRST variations of the fermions in (\ref{eq:conctopgrav}) and (\ref{eq:concSgauge}) we obtained  equations for supersymmetric backgrounds
\bea
&& \mathcal{L}_\gamma\, g_{\mu\nu}=0\nn\\
&&D\, \phi - i_\gamma \, (F) =0 
\label{con:eqequationone}
\eea
These equations lead to the $\gamma$-equivariant cohomology equations (\ref{eq:chernequivarianteq}), for the equivariant
Chern classes of the supergravity Yang-Mills gauge bundle. 

From the same BRST algebra (\ref{con:brstalgebra}) we derived additional equations, valid on the localization locus,  for certain  {\it gauge invariant} scalar ghost bilinears $\varphi$.  These equations also take the form of  $\gamma$-equivariant cohomology equations
\bea
d\, \varphi - i_\gamma(T^{(2)})=0
\label{con:eqequationtwo}
\eea
The scalar gauge invariant ghost bilinears $\varphi$ can be constructed only for certain extended supergravity with specific Yang-Mills groups. The corresponding bosonic 2-forms $T^{(2)}$ are model dependent. In the last 
section of this paper we have analyzed in detail supergravity with  $N=(2,2) $ in $d=2$
for which two such bilinears $\varphi$ can be constructed. We have seen how the  $\gamma$-equivariant cohomological equations (\ref{con:eqequationone}) and (\ref{con:eqequationtwo}) are related to each other, thus providing an {\it a priori} explanation of the results presented in \cite{Bae:2015eoa}.

Another theory for which two gauge invariant scalar bilinears $\varphi$ can be constructed is $N=2$ $d=4$ supergravity. The study of the cohomological  equations (\ref{con:eqequationtwo}) and (\ref{con:eqequationone}) for this theory might lead to the classification  of localizing backgrounds for this theory --- a long standing problem to which we hope to come back in the future.

We found two kinds of topological multiplets inside supergravity: the ``universal'' ones, whose 2-form components are the curvatures 
and whose 0-form components are the model dependent scalars $\phi$ which appear in the BRST algebras; and the gauge invariant scalars
multiplets, which exist only for certain supergravities, whose 0-form components are made entirely of commuting ghosts and
whose 2-forms components are model dependent. For $N=(2,2) $ in $d=2$ supergravity we found compact quadratic relations, ultimately descending from the Fierz identities, between these
topological multiplets. It would be interesting to understand
if generalizations of these relations exists for other models, notably for $N=2$ $d=4$ supergravity: beyond their application
to localization these might give important informations about the dynamics of supergravity itself.

\section*{Acknowledgments}

We thank S.~Ferrara, S.~Murthy and J.~Winding  for useful discussions.

CI thanks the Theory Group of KIAS, Seoul and of LPTHE, of the Sorbonne University and CNRS Paris; DR thanks INFN Genoa and the Physics Department of the University of Genoa for their hospitality and financial support during part of this work. 

The work of CI was supported in part by  INFN, by Genoa University Research Projects, F.R.A. 2015. 
\appendix
\section{Equivariant cohomology in dimension 2}
\label{app:eqcohomology2}

In this Appendix we collect some facts   regarding equivariant
forms of degree 2 in 2 dimensions, which are needed to derive the relations (\ref{N2d2compositeness}) between  the curvature polyforms of $N=(2,2)$ in $d=2$ supergravity and the gauge invariant polyforms $\mathbb{H}_i$, with $i=1,2$.

Degree 2 $\gamma$-equivariant forms $\Phi$ are polyforms 
\bea
\Phi = \Phi^{(2)}+ \Phi^{(0)} 
\eea
with
\bea
\mathcal{D}_\gamma \,\Phi=0
\label{u1eq}
\eea
where $\mathcal{D}_\gamma= d- i_\gamma$   is the $\gamma$-equivariant exterior derivative associated to the Killing vector $\gamma^\mu$ that we introduced in (\ref{eq:coboundaryDgamma}).  In dimension 2,  the 0-form and the 2-form component of a $\gamma$-equivariant form of degree 2 are
related by
\bea
\Phi^{(2)} = k \, d \, \Phi^{(0)}
\label{Phi2Phi0eq}
\eea
where the 1-form $k$  is the ``inverse'' of the Killing vector $\gamma^\mu$\footnote{Since $\gamma^\mu$ may have zeros, $k$ may have poles. However, one can show that $\Phi^{(2)}$ defined in (\ref{Phi2Phi0eq}) is regular when $\Phi^{(0)}$ satisfies the equivariance equation (\ref{u1eq}).}
 \bea
&&  k \equiv \frac{g_{\mu\nu}\, \gamma^\nu}{\gamma^2}\, dx^\mu \qquad
 \gamma^2 \equiv g_{\mu\nu}\ \gamma^\mu\, \gamma^\nu\nn\\
  && i_\gamma(k) =1
 \eea
The 2-form  $\Phi^{(2)}$  in (\ref{Phi2Phi0eq}) depends on the choice of a metric $g_{\mu\nu}$. However, different metrics with the same Killing
vector $\gamma^\mu$ give rise to  1-forms $k$ which differ by 1-forms $\delta k$ which are both $d$ and $i_\gamma$ 
closed
\bea
i_\gamma( \delta k) = d\, (\delta k) =0
\eea
Therefore polyforms with  (\ref{Phi2Phi0eq}) corresponding to different metrics are cohomologous
\bea
\delta \,\Phi^{(2)} = d\,  (\delta k \, \Phi^{(0)}) \qquad \delta \,\Phi^{(0)} = i_\gamma (\delta k \, \Phi^{(0)})=0
\eea
  
Let us also observe that  the product of two 
$\gamma$-equivariant forms of degree 2 in 2 dimensions is again a $\gamma$-equivariant form of degree 2,  i. e. the set of  $\gamma$-equivariant forms of degree $2$ in $2$ dimensions has a ring structure:
\bea
&&\Phi_1\, \Phi_2 =  \Phi^{(2)}_1\,\Phi^{(0)}_2+ \Phi^{(2)}_0\,\Phi^{(0)}_1+\Phi^{(0)}_1\, \Phi^{(0)}_2\nn\\
&&  \mathcal{D}_\gamma \,(\Phi_1\, \Phi_2) =0
\label{Phiproduct}
 \eea
Moreover, given a metric, one can introduce a linear operation $L$ which when acting on a $\gamma$-equivariant form of degree 2
gives another $\gamma$-equivariant form of degree 2:
\bea
&& L\,(\Phi) = \star \big[D_\mu\, \bigl(\frac{1}{\gamma^2}\, D^\mu\,\Phi^{(0)}\bigr)\bigr]+ \star \Phi^{(2)}\equiv\nn\\
&& \qquad \equiv  \star (\Delta_\gamma\,\Phi^{(0)})+ \star \Phi^{(2)}\ \nn\\
&& \mathcal{D}_\gamma \,L\,(\Phi)=0
\label{app:Loperator}
 \eea
where we introduced the $\gamma$-dependent Laplacian
 \bea
 \Delta_\gamma\,\Phi^{(0)}\equiv D_\mu\, \bigl(\frac{1}{\gamma^2}\, D^\mu\,\Phi^{(0)}\bigr)
 \eea
 for 0-forms\footnote{Since $\gamma^\mu$ may in general have zeros, the action of  $ \Delta_\gamma$
 on a generic 0-form is not always well-defined. However it can be shown that $ \Delta_\gamma\, \Phi^{(0)}$ is  regular
 when $\Phi^{(0)}$ is the 0-form component of a $\gamma$-equivariant form of degree 2.}.  $L$ acts as a derivative with respect to the product (\ref{Phiproduct})
 \bea
L \,\bigl(\Phi_1\, \Phi_2 \bigr) = L\, (\Phi_1)\, \Phi_2 + \Phi_1\, L\, (\Phi_2)
 \eea
 It should be kept in mind the $L$ does not map  $\gamma$-equivariant forms to cohomologous ones.
 
  Let us observe the cohomology class of a $\gamma$-equivariant form $\Phi$ of degree 2 on the 2-dimensional sphere is parametrized by the values of the polyform at the two zeros  $p_\pm$ of the Killing vector $\gamma^\mu$:
 \bea
 \Phi(p_\pm) = \Phi^{(0)}(p_\pm)
\eea
The cohomological invariant obtained by evaluating $\Phi$ on the sphere is related to  $ \Phi(p_\pm) $ by the localization
formula
 \bea
 \int_{S^2} \, \Phi= \int_{S^2} \, \Phi^{(2)} =  \Phi^{(0)}(p_+)- \Phi^{(0)}(p_-)
 \eea
 Let us introduce the equivariantly closed polyform
 \bea
 && \hat{\mathbb{R}} = - \eta^{ij}\, \Phi_i \,\Phi_j = 
 -\bigl(\gamma^2  + k\, d \gamma^2 \bigr)
 \label{Fierzeq}
 \eea
 where we used the Fierz identity $\varphi_1^2- \varphi_2^2= \gamma^2$. Note that
 \bea
 \mathbb{R} = L\, (\hat{\mathbb{R}})=  \phi_{\mathrm{Lor}} + R^{(2)}
\label{PhiLhatPhiL}
\eea
From (\ref{Fierzeq}) one obtain the following expression for the $\gamma$-equivariant extension of curvature 2-form:
\bea
&&  \mathbb{R} = \frac 1 2 \sqrt{g}\, \epsilon_{\mu\nu}\, D^\mu\, \gamma^\nu + R^{(2)}=\nn\\
&&\qquad = i_\gamma\bigl( \star\, \frac{ d\, \gamma^2}{\gamma^2} \bigr)- d\, \star\,  \bigl(\frac{ d\, \gamma^2}{\gamma^2} \bigr)
\eea
This equivariant form depends on the metric and the corresponding Killing vector. Under a variation of the metric $\delta\,g_{\mu\nu}$,
keeping the Killing vector fixed, it varies by cohomologically trivial terms
\bea
&&\delta R^{(2)}= d\, \delta\omega^{(1)} \qquad \delta  \phi_{\mathrm{Lor}} = i_\gamma( \delta\, \omega^{(1)})\nn\\
&&\qquad \delta\omega^{(1)}= \delta\bigl( \star\,  \frac{d\gamma^2}{\gamma^2}\bigr)
\eea
In particular
\bea
 \delta \mathbb{R} (p_\pm)=0
 \eea
 Therefore the $\gamma$-equivariant cohomology class of $\mathbb{R}$ is independent of the metric and can be computed, for
 example, using the round metric:
 \bea
   \mathbb{R}(p_\pm)= \mp 1
   \label{Rchern}
  \eea

\newpage
\bibliography{ir}
\bibliographystyle{ir}
\end{document}